\documentclass[a4paper,11pt]{article}

\usepackage{amsmath,amssymb}
\usepackage{bm}
\usepackage{upgreek}
\usepackage{textcomp}
\usepackage{url}
\usepackage{rotating}
\usepackage{gensymb}
\usepackage{natbib}
\usepackage[english]{babel}
\usepackage[T1]{fontenc}
\usepackage{float}
\usepackage[top=1in, bottom=1.25in, left=1.25in, right=1.25in]{geometry}
\usepackage{adjustbox}

\title{Gradient-based deterministic inversion of geophysical data with Generative Adversarial Networks: is it feasible?}

\author{Eric Laloy\thanks{Belgian Nuclear Research Centre, Email: {\tt elaloy@sckcen.be}}, Niklas Linde\thanks{University of Lausanne}, Cyprien Ruffino\thanks{INSA-Rouen}, Romain H\'erault$^{\ddagger}$, Gilles Gasso$^{\ddagger}$, \\
	and Diederik Jacques$^{*}$}
\begin{document}

\maketitle

\begin{abstract}
	
Global probabilistic inversion within the latent space learned by a Generative Adversarial Network (GAN) has been recently demonstrated. Compared to inversion on the original model space, using the latent space of a trained GAN can offer the following benefits: (1) the generated model proposals are geostatistically consistent with the prescribed prior training image (TI), and (2) the parameter space is reduced by orders of magnitude compared to the original model space. Nevertheless, exploring the learned latent space by state-of-the-art Markov chain Monte Carlo (MCMC) methods may still require a large computational effort. As an alternative, parameters in this latent space could possibly be optimized with much less computationally expensive gradient-based methods. This study shows that due to the typically highly nonlinear relationship between the latent space and the associated output space of a GAN, gradient-based deterministic inversion may fail even when considering a linear forward physical model. We tested two deterministic inversion approaches: a quasi-Newton gradient descent using the Adam algorithm and a Gauss-Newton (GN) method that makes use of the Jacobian matrix calculated by finite-differencing. For a channelized binary TI and a synthetic linear crosshole ground penetrating radar (GPR) tomography problem involving 576 measurements with low noise, we observe that when allowing for a total of 10,000 iterations only 13\% of the gradient descent trials locate a solution that has the required data misfit. The tested GN inversion was unable to recover a solution with the appropriate data misfit. Our results suggest that deterministic inversion performance strongly depends on the inversion approach, starting model, true reference model, number of iterations and noise realization. In contrast, computationally-expensive probabilistic global optimization based on differential evolution always finds an appropriate solution.
	
\end{abstract}

\section{Introduction and Scope}
\label{intro}

\citet[][]{Laloy2018} recently proposed to use generative adversarial networks (GANs), a game changer data generation algorithm \citep[e.g.,][]{Goodfellow2014,Goodfellow2016}, to define a low-dimensional parameterization encoding complex geologic prior models, thereby allowing efficient and accurate geostatistical inversion with Markov chain Monte Carlo (MCMC) methods \citep[][]{Laloy2018}. GANs have permitted impressive advancements for a wide range of applications such as image and texture synthesis, image-to-image translation and super-resolution \citep[][]{Creswell2017}. In the near future, we expect to witness a dramatic increase in development and use of GAN-inspired algorithms for geostatistical simulation \citep[e.g.,][]{Mosser2017,Laloy2018} and inversion \citep[e.g.,][]{Laloy2018,Mosser2018,Richardson2018}. A key element of the GAN approach is that the dimensions of the learned low-dimensional or ``latent'' space are independent from each other and have a known probability distribution, typically either a uniform or a standard Gaussian distribution. \citet[][]{Laloy2018} have shown that inversions based on such parameterizations work well for global probabilistic inference of complex binary 2D and 3D prior subsurface models. However, exploring the GAN-derived latent space with state-of-the-art MCMC sampling \citep[][]{Vrugt2009, Laloy-Vrugt2012} still necessitates tens of thousands (or more) forward evaluations \citep[][]{Laloy2018}. Such a computational expense can be prohibitive when using computationally-demanding forward solvers encountered in the geosciences. However, the latent space might also lend itself to a conventional deterministic gradient-based local search, which is often much more computationally frugal than global and probabilistic inversion.

In the geosciences, gradient-based deterministic inversion methods can be roughly divided into (1) methods that make use of the so-called sensitivity or Jacobian matrix, $\textbf{J}$, that is, the matrix of derivatives of the forward model outputs, $F\left(\textbf{m}\right)$, with respect to the model parameters, $\textbf{m}$ (e.g., physical properties of the model grid blocks): $J_{i,j} = \displaystyle\frac{\delta F\left(\textbf{m}\right)_i}{\delta m_j}$, and (2) methods that only require the gradient vector of the misfit (that is, objective or loss) function, $\mathcal{L}$, with respect to $\textbf{m}$, $\nabla \mathcal{L}$ with $\nabla \mathcal{L}_{j} = \displaystyle\frac{\delta \mathcal{L}}{\delta m_j}$, such as the steepest descent method. Although $\nabla \mathcal{L}$ is often cheaper to calculate than $\textbf{J}$, it is generally understood that methods relying on $\textbf{J}$ are more effective and robust than those based on $\nabla \mathcal{L}$ because $\textbf{J}$ can be used to (implicitely) approximate the Hessian matrix, $\textbf{H}$ with $H_{i,j} = \displaystyle\frac{\delta J_{i,j}}{\delta m_j}$, which contains information about the local curvature of the misfit function \citep{Aster2012}.

Complex spatial prior models are usually represented by a so-called training image (TI). A TI is a large gridded 2D or 3D unconditional representation of the expected target spatial field that can be either continuous or categorical (e.g., geologic facies image) and is typically used to guide geostatistical simulation by multiple-point statistics (MPS) algorithms \citep[e.g,][]{Strebelle2002,Mariethoz2010}. When using a GAN \citep[see, e.g.,][for details about the GAN architecture and training]{Goodfellow2016} to encode the prescribed TI (or, alternatively, MPS realizations from it), a GAN-based model realization is produced by feeding the so-called generator, $G$, with a latent vector, $\textbf{z}$: $\textbf{m} = G\left(\textbf{z}\right)$, and the inversion is performed within the latent space $p\left(\textbf{z}\right)$. For gradient-based inversion, the required Jacobian is then the matrix of derivatives of $F\left[G\left(\textbf{z}\right)\right]$ with respect to $\textbf{z}$: $\textbf{J}^{\rm z}$ with $J^{\rm z}_{i,j} = \displaystyle\frac{\delta F\left[G\left(\textbf{z}\right)\right]_i}{\delta z_j}$. In principle, the derivatives of $G\left(\textbf{z}\right)$ with respect to $\textbf{z}$, $\textbf{V}$ with $V_{i,j} = \displaystyle\frac{\delta G\left(\textbf{z}\right)_i}{\delta z_j}$, can be directly computed by autodifferentiation \citep[][]{theano2016, tensorflow2016, pytorch2017}. When $\textbf{J}$ is available, for instance through the use of an adjoint model, $\textbf{J}^{\rm z}$ can be obtained as $\textbf{J}^{\rm z} = \textbf{J}\textbf{V}$. In our experience, calculating each element of $\textbf{V}$ using the reverse-mode autodifferentiation engine that equips current deep learning (DL) libraries such as TensorFlow \citep{tensorflow2016} and PyTorch \citep{pytorch2017} can be slow, especially for large $\textbf{m}$ and when run on a CPU. For instance, for the PyTorch implementation of the spatial GAN \citep{Jetchev2016} used herein it takes about 5 minutes to construct the $\textbf{V}$ array on a last generation Intel\textsuperscript{\textregistered} i7 CPU and about 1 minute using a NVIDIA Quadro M2000M GPU, when $\textbf{z}$ is 15-dimensional and $G\left(\textbf{z}\right)$ is of size 129 $\times$ 65. More importantly, generating high-quality categorical geologic structures with a GAN may require postprocessing of the realizations through filtering and/or thresholding \citep[e.g.,][]{Mosser2017, Laloy2018}. If not differentiable, this postprocessing, $pp\left(\cdot\right)$, decouples the $\textbf{V}$ array computed by the DL code from the actual realization, $pp\left[G\left(\textbf{z}\right)\right]$.

Another solution that alleviates the need of having an adjoint model and is not perturbed by postprocessing operations consists of estimating $\textbf{J}^{\rm z}$ directly using a finite-difference approximation. This incurs a relatively low computational cost since it requires $2N_z + 1$ forward runs to evaluate $\textbf{J}^{\rm z}$ with a central difference scheme, with $N_z$ the length of the $\textbf{z}$ vector, which by construction is low-dimensional (say two orders of magnitude less than $\textbf{m}$).

Lastly, when using the gradient vector, $\nabla \mathcal{L}$ with respect to $\textbf{z}$, $\nabla \mathcal{L}\left(\textbf{z}\right)$, instead of the Jacobian, $\textbf{J}^{\rm z}$, $\nabla \mathcal{L}\left(\textbf{z}\right)$ and the trained $G\left(\textbf{z}\right)$ can be integrated within a single DL computational graph \citep[][]{Richardson2018,Mosser2018}. This is possible because $G\left(\textbf{z}\right)$ is fully differentiable end-to-end and DL libraries are equipped with advanced gradient descent algorithms such as Adam \citep[][]{Kingma-Ba2015}. If $F\left[G\left(\textbf{z}\right)\right]$ is fully differentiable then the DL library can compute $\nabla \mathcal{L}\left(\textbf{z}\right)$ internally by autodifferentiation. This is for instance the case when $F\left(\cdot\right)$ is linear and the continuous $G\left(\textbf{z}\right)$ realizations do not need postprocessing such as thresholding. As shown by \citet[][]{Richardson2018}, when $pp\left[G\left(\textbf{z}\right)\right]$ or $F\left(\cdot\right)$ is not differentiable $\nabla \mathcal{L}\left(\textbf{m}\right)$ can however be externally computed and integrated within the DL graph. In this work, we used an adjoint model giving $\nabla \mathcal{L}\left(\textbf{m}\right)$ and backpropagated $\nabla \mathcal{L}\left(\textbf{z}\right) = \nabla \mathcal{L}\left(\textbf{m}\right)\textbf{V}$ where the $\nabla \mathcal{L}\left(\textbf{m}\right)\textbf{V}$ product can be efficiently computed within a single backward pass without explicitely calculating the $\textbf{V}$ array (see our companion code).

No matter whether $\nabla \mathcal{L}\left(\textbf{z}\right)$ or $\textbf{J}^{\rm z}$ is used or regardless of how they are calculated, an important open question is to what extent the nonlinearity in the $G\left(\textbf{z}\right)$ transform adversely affects the convergence of gradient-based deterministic inversion performed in the latent space $p\left(\textbf{z}\right)$. Working with $p\left(\textbf{z}\right)$ rather the original model space $p\left(\textbf{m}\right)$ conveniently ensures that (1) any generated model honors the prior TI and (2) the parameter dimensionality is reduced by orders of magnitude compared to traditional deterministic inversion \citep[e.g.,][]{deGroot-Hedlin-Constable1990}. Nevertheless, the non-linear relationship between $G\left(\textbf{z}\right)$ and $\textbf{z}$ adds significant nonlinearities to the inverse problem, in addition to those of the forward model, $F\left(\textbf{m}\right)$. In this work, we consider a best case scenario and use a linear tomography problem based on ground penetrating radar (GPR) data to show that even if $F\left(\textbf{m}\right)$ is linear with respect to $\textbf{m}$, the nonlinearity in $G\left(\textbf{z}\right)$ may frequently affect the performance of gradient-based deterministic inversions. In contrast, probabilistic global optimization using the DREAM$_{\left(\rm ZS\right)}$ code \citep[][]{Vrugt2009, Laloy-Vrugt2012} is found to work well for the considered problem, although at a rather high computational cost. The overall motivation of this work is to raise awareness that although model parameterizations based on GAN algorithms can be very powerful in representing prior information and reducing model dimensionality, the resulting inverse problems may become too non-linear to enable adequate convergence of deterministic gradient-based inversions even when the forward solver itself is only weakly-nonlinear or even linear.

The remainder of this paper is organized as follows. Section \ref{related_work} summarizes related work before section \ref{methods} briefly describes our used GAN and considered inversion algorithms, and details the considered inverse problem. This is followed by section \ref{results} which presents our inversion results. In section \ref{discussion}, we discuss our main findings and outline possible future developments. Finally, section \ref{conclusion} provides a short conclusion.

\section{Related work}
\label{related_work}

To the best of our knowledge, \citet{Laloy2018} were the first to introduce and demonstrate the idea of using the latent space learned by a GAN to perform probabilistic inversion of hydrologic or geophysical data. \citet{Richardson2018} recently proposed to embed $\nabla \mathcal{L}\left(\textbf{m}\right)$ and a generator, $G\left(\textbf{z}\right)$, trained beforehand within a single (fully differentiable) DL computational graph to deterministically solve a 2D seismic full-waveform inversion problem by making use of the stochastic gradient descent algorithms implemented in DL libraries. \citet{Mosser2018} recently used a trained generator for solving a 2D seismic inversion problem within a probabilistic framework. They combined the estimation of $\nabla \mathcal{L}\left(\textbf{m}\right)$ obtained from an adjoint model together with $G\left(\textbf{z}\right)$ and the DL-based calculation of $\textbf{V}$ within the same (fully differentiable) computational graph to perform Metropolis-adjusted Langevin MCMC sampling \citep[e.g.,][]{Roberts-Rosenthal1998}. Outside of the geosciences, \citet{Creswell-Bharath2016} and \citet{Lipton-Tripathi2017} inverted full images produced by a trained generator using usual benchmarks in computer vision and studied to what extent latent vectors can be recovered. In addition, \citet{Bora2017} showed that gradient descent within the latent space of a GAN works overall well to recover an image from a set of measurements obtained by applying a linear operator to the true image. Conceptually, the \citet{Bora2017} study has a lot in commons with ours. However, Bora and coworkers consider a totally different training set than us: the CelebA dataset which consists of face images of celebrities. In section \ref{discussion}, we compare our findings to those of \citet{Bora2017}.

\section{Methods}
\label{methods}

\subsection{Generative Adversarial Networks}
\label{meth_gan}

\begin{figure}[H]
	\noindent\hspace{-2.6cm}\includegraphics[width=45pc]{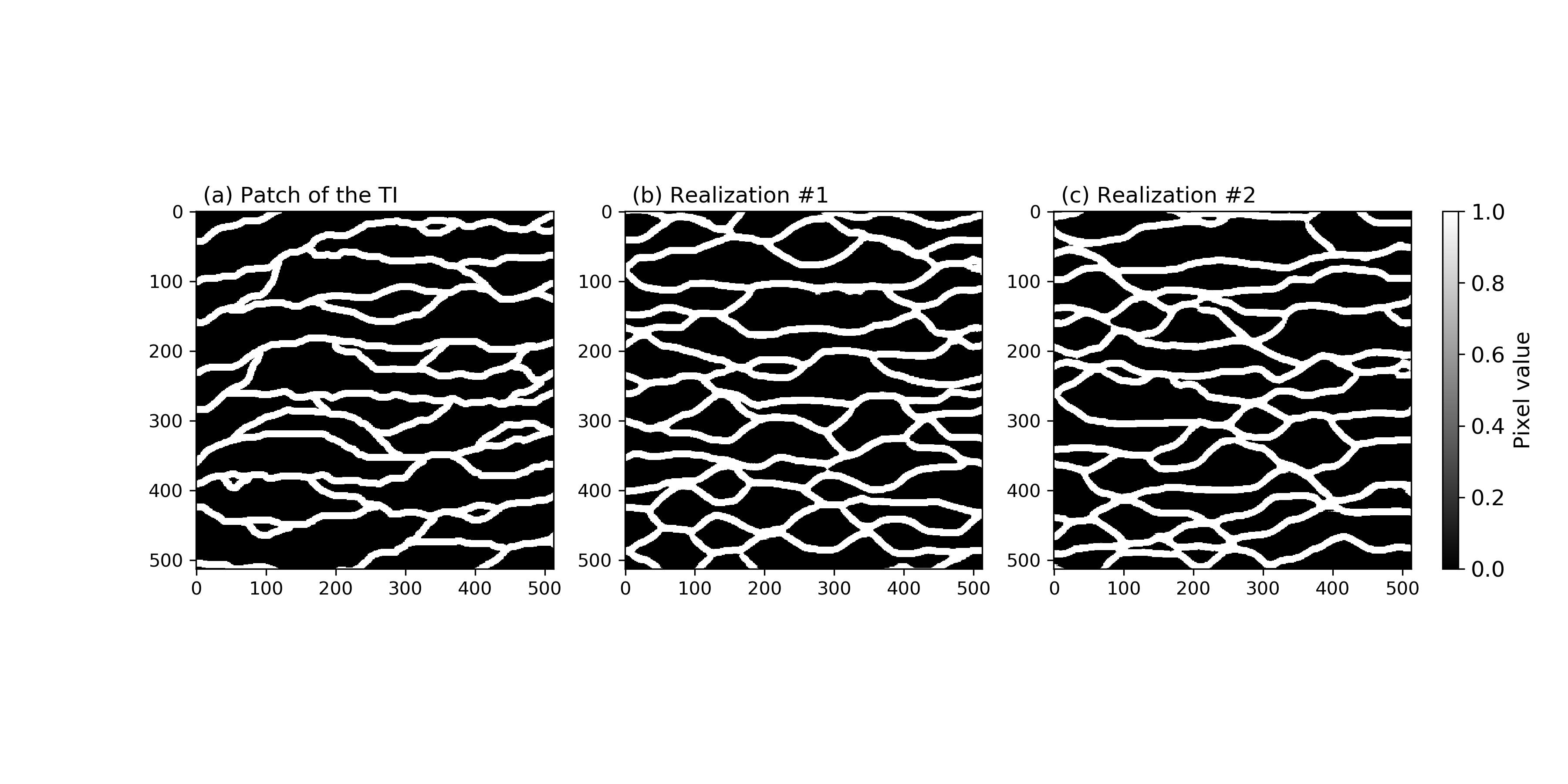}
	\vspace*{-2.4cm}
	\caption{(a) Fraction of size $513 \times 513$ of the used $2500 \times 2500$ binary TI and (b) - (c) randomly chosen $513 \times 513$ realizations derived by our SGAN. Each continuous realization is generated by sampling 289 random numbers from a uniform distribution, $U\left(-1,1\right)$, without any postprocessing.}
	\label{fig1}
\end{figure}

We use a PyTorch implementation of the spatial GAN \citep[SGAN,][]{Jetchev2016} used by \citet{Laloy2018} to generate realizations from the 2D channelized aquifer training image (TI) depicted in Figure \ref{fig1}a. Our method is not limited to such channel images, but this example is selected to be similar to benchmark images that are typically used to test MPS algorithms. For brevity, we refer the reader to \citet{Laloy2018} for a description of the basic network architecture and training procedure, and only provide below the main principles behind the used SGAN.

The building blocks of a GAN are the generator and discriminator. The generator, $G\left(\textbf{z}\right)$, is fed with a low-dimensional latent vector, $\textbf{z}$,  and produces a model realization, $G\left(\textbf{z}\right) = \textbf{m}$, whose spatial statstics match those found in the TI provided that the GAN training performed beforehand (see below) was successful. The $\textbf{z}$ vector is trained to obey either a standard normal distribution, $\textbf{z} \propto N(\textbf{0},\textbf{I})$, or a truncated uniform distribution, $\textbf{z} \propto U(\textbf{-1},\textbf{1})$, and the $z$ variables are independent of each other. For image generation, the discriminator component only serves for training $G\left(\textbf{z}\right)$ as detailed below.

Due to its purely convolutional nature \citep[see, e.g.,][for more details on convolutions and convolutional layers in deep neural networks]{Goodfellow2016}, for an ergodic TI our SGAN can be trained at relatively low computational cost using a small realization domain, before being used to generate arbitrarily large realizations. The latent space of our SGAN has a spatial structure, with $\textbf{z}$ being reshaped into an $m \times n \times o \times q$ array $\textbf{Z}$ for the 3D case. As detailed in \citet{Laloy2018}, the $m$, $n$ and $o$ dimensions are directly related to the three spatial dimensions while $q$, which we set to 1 in this work, is an extra dimension that can encode additional information about the data representation \citep[see geostatistical simulation case study 2 in][]{Laloy2018}. For a square ($w \times h$) or cubic ($w \times h \times l$) generation domain, the relationship  between $z_{\rm x}= m = n = o$ and $m_{\rm x} = w = h = l$ is given by
	\begin{equation}
	m_{\rm x} = \left[z_{\rm x} - 1\right]2^{dp} + 1,
	\label{sgan1}
	\end{equation}
where $dp$ is the number of stacked convolutional layers in $G\left(\textbf{z}\right)$. This allows for a rather strong dimensionality reduction. For example, when $z_{\rm x} = 5$ and $dp = 5$, we have $m_{\rm x} = 129$. We refer to \citet{Jetchev2016} for more details on the (2D) SGAN architecture.

In a GAN, the generator and discriminator are trained (or ``learned") simultaneously with opposing goals. The discriminator, $D\left(\textbf{m}\right)$, is fed with samples from the ``real" training set, which from now on will be referred to as $\textbf{m}_{\rm true}$ with distribution $p_{\rm data}\left(\textbf{m}\right)$, and ``fake" samples (i.e., realizations) created by the generator: $\textbf{m} = G\left(\textbf{z}\right)$. In our case, the real samples $\textbf{m}_{\rm true}$ are a set of patches randomly cut from the TI. The discriminator tries to distinguish between $\textbf{m}_{\rm true}$ and $\textbf{m}$ by computing, for each received sample, the probability that it belongs to $p_{\rm data}\left(\textbf{m}_{\rm true}\right)$. Conversely, the generator, $G\left(\textbf{z}\right)$, aims at fooling $D\left(\cdot\right)$ into labeling $\textbf{m}$ as a sample from $p_{\rm data}\left(\textbf{m}_{\rm true}\right)$ \citep{Goodfellow2014}. This translates into the following minimization-maximization problem
	\begin{equation}
	\min_{G\left(\cdot\right)} \max_{D\left(\cdot\right)}\left\{\mathbb{E}_{\textbf{m}_{\rm true}\sim p_{\rm data}\left(\textbf{m}_{\rm true}\right)}\left[\log\left(D\left(\textbf{m}_{\rm true}\right)\right)\right] +\mathbb{E}_{\textbf{z}\sim p_{\textbf{z}}\left(\textbf{z}\right)}\left[\log\left(1-D\left(G\left(\textbf{z}\right)\right)\right)\right]\right\},
	\label{sgan2}
	\end{equation}

Compared to the SGAN code used by \citet{Laloy2018}, we replaced batch normalization by instance normalization and added two transposed dilated convolutional layers to the generator, before its output (see section \ref{appendix_archi}). Upon selection of the appropriate training epoch (see below), this removes the need to post-process the realizations by median filtering to eliminate small impurities, thereby leaving us with thresholding as the unique postprocessing operation. The realizations produced by our trained $G\left(\textbf{z}\right)$ are continuous on the $\left[0,1\right]$ range and they are converted into binary images by thresholding at the 0.5 level.

Training was performed on a GPU Tesla K40 for 50 epochs with 64 mini-batches of 100 training samples per epoch. Epoch 36 was deemed to produce the best realizations. Figures \ref{fig1}b-c show two (randomly chosen) realizations generated by the trained SGAN before thresholding, $tr\left[G\left(\textbf{z}\right)\right]$. These 513 $\times$ 513 realization are obtained by sampling a 289-dimensional $\textbf{z}$ vector from $p\left(\textbf{z}\right) \propto U(\textbf{-1},\textbf{1})$ \citep[see][for details and a performance comparison with a popular MPS algorithm]{Laloy2018}. Even if these images are continuous, they appear almost categorical in this representation, which also suggests that the subsequent thresholding is a relatively mild operation.

\subsection{Gradient-Based Deterministic Inversion}
\label{meth_gradinv}

A common representation of the forward problem is
	\begin{equation}
	\textbf{d} = F\left(\textbf{m}\right) + \textbf{e},
	\label{gbinv1}
	\end{equation}
where $\textbf{d} = \left(d_1, \ldots, d_N \right) \in \mathbb{R}^N, N \geq 1$ is the number of measurement data, $F\left(\textbf{m}\right)$ denotes a deterministic forward model with parameters or model $\textbf{m}$ and the noise term $\textbf{e}$ lumps all sources of errors. If the probability distribution of $\textbf{e}$ can be assumed to be zero-mean Gaussian with covariance matrix $\textbf{C}_{\rm e \it}$, and $\textbf{m}$ is assigned a multiGaussian prior distribution, then the optimal model, $\hat{\textbf{m}}$, minimizes the following objective or loss function, $\mathcal{L}\left(\textbf{m}\right)$
	\begin{equation}
	\mathcal{L}\left(\textbf{m}\right) = \Phi_D + \lambda\Phi_M,
	\label{gbinv2}
	\end{equation}
	\begin{equation}
	\Phi_D = \frac{1}{2}\left[F\left(\textbf{m}\right) - \textbf{d}\right]^{\rm T \it}\textbf{C}_{\rm e \it}^{-1}\left[F\left(\textbf{m}\right) - \textbf{d}\right],
	\label{gbinv3}
	\end{equation}
	\begin{equation}
	\Phi_M = \frac{1}{2} \left[\textbf{m} - \textbf{m}_{\rm prior \it}\right]^{\rm T \it}\textbf{C}_{\rm m \it}^{-1}\left[\textbf{m} - \textbf{m}_{\rm prior \it}\right],
	\label{gbinv4}
	\end{equation}
where $\textbf{m}_{\rm prior \it}$ and $\textbf{C}_{\rm m \it}$ signify the a priori model and its covariance matrix, respectively, and where $\lambda$ weights the influence of honoring prior Gaussian information about $\textbf{m}$ on $\mathcal{L}\left(\textbf{m}\right)$. In case of a linear relationship between $\textbf{d}$ and $F\left(\textbf{m}\right)$, $F\left(\textbf{m}\right)$ reduces to the product $\textbf{F}\textbf{m}$. In practice, it is common to replace $\textbf{C}_{\rm m \it}$ by a regularization operator, $\textbf{L}$. This leads to the joint minimization of  the misfit, $\Phi_D$, and $\Phi_M = ||\textbf{L}\left(\textbf{m}-\textbf{m}_{\rm prior \it}\right)||_2^2$

	\begin{equation}
	\mathcal{L}\left(\textbf{m}\right) = \frac{1}{2}\left[F\left(\textbf{m}\right) - \textbf{d}\right]^{\rm T \it}\textbf{C}_{\rm e \it}^{-1}\left[F\left(\textbf{m}\right) - \textbf{d}\right] + \lambda\left[\textbf{m} - \textbf{m}_{\rm prior \it}\right]^{\rm T \it}\textbf{L}^{\rm T \it}\textbf{L}\left[\textbf{m} - \textbf{m}_{\rm prior \it}\right].
	\label{gbinv5}
	\end{equation}

The popular Gauss-Newton (GN) method for minimizing equation (\ref{gbinv5}) iteratively updates $\textbf{m}$ until either the target data misfit or a maximum number of iterations has been reached using the following update mechanism

	\begin{equation}
	\textbf{m}_{k+1} = \textbf{m}_k + \Delta\textbf{m} = \left(\textbf{J}_k^{\rm T \it}\textbf{C}_{\rm e \it}^{-1}\textbf{J}_k^{\rm T \it}+\lambda \textbf{L}^{\rm T \it}\textbf{L}\right)^{-1}\textbf{J}_k^{\rm T \it}\textbf{C}_{\rm e \it}^{-1}\hat{\textbf{d}_k}+\textbf{m}_{\rm prior \it},
	\label{gbinv6}
	\end{equation}
with
	\begin{equation}
	\hat{\textbf{d}_k} = \textbf{d} - F\left(\textbf{m}_k\right) + \textbf{J}_k\left(\textbf{m} - \textbf{m}_{\rm prior \it}\right),
	\label{gbinv7}
	\end{equation}
where $k$ denotes the iteration number. In the context of our GAN-based dimensionality reduction, we learn a $p\left(\textbf{z}\right) = U\left(\textbf{-1},\textbf{1}\right)$ for the latent vector $\textbf{z}$. Limited testing with learning a standard normal distribution, $N\left(\textbf{0},\textbf{1}\right)$, is found to provide good results but the generated realizations are nevertheless of slightly lower quality than when using $U\left(\textbf{-1},\textbf{1}\right)$. Fortunately, when the components of a random vector, $\textbf{x}$ are independent, they can be converted into standard normal variables using the iso-probabilistic transform: labeling as $\uptau_{\rm x \it}\left(x_i\right)$ ($\Psi\left(\xi_i\right)$) the cumulative density function (CDF) of $x_i$ (a standard normal variable $\xi_i$), the direct and inverse transform are given by
	\begin{equation}
	\xi_i = \Psi^{-1}\left[\uptau_{\rm x \it}\left(x_i\right)\right] \ \ \ \rm{and} \it \ \ \ x_i = \uptau_{\rm x \it}^{-1}\left[\Psi\left(\xi_i\right)\right].
	\label{gbinv8}
	\end{equation}
To perform the inversion within the GAN latent space, we therefore place a standard normal uncorrelated prior (so-called damping regularization) on a vector $\textbf{z}_{\rm SN \it}$ of the same size as $\textbf{z}$ and  achieve GN updates of $\textbf{z}_{\rm SN \it}$ while converting each proposed $\textbf{z}_{\rm SN \it}$ into the corresponding $\textbf{z} \propto U\left(\textbf{-1},\textbf{1}\right)$ with equation (\ref{gbinv8}) before creating the resulting $G\left(\textbf{z}\right)$ realization. Under the standard normal prior on $\textbf{z}_{\rm SN \it}$, the GN update simplifies to
	\begin{equation}
	\textbf{z}_{\rm SN \it, k+1} = \left(\textbf{J}_{\rm z, \it k}^{\rm T \it}\textbf{C}_{\rm e \it}^{-1}\textbf{J}_{\rm z, \it k}^{\rm T \it}+\lambda \textbf{I}\right)^{-1}\textbf{J}_{\rm z, \it k}^{\rm T \it}\textbf{C}_{\rm e \it}^{-1}\left[\textbf{d} - F\left(G\left( \uptau_{\rm z \it}^{-1}\left[\Psi\left(\textbf{z}_{\rm SN \it, k}\right)\right] \right)\right) + \textbf{J}_{\rm z, \it k}\textbf{z}_{\rm SN \it, k} \right]
	\label{gbinv9}
	\end{equation}
where $\textbf{J}_{\rm z, \it k} = \textbf{J}^{\rm z_{\rm SN}}_k$ is the Jacobian matrix of $\textbf{z}_{\rm SN \it, k}$, and $\textbf{I}$ is the identity matrix. In this work, we simply set $\lambda = 1$ as adjusting $\lambda$ dynamically did not show any advantages over fixing $\lambda = 1$.

When only the gradient of the objective or loss function with respect to the model parameters, $\nabla \mathcal{L}\left(\textbf{m}\right)$, is used, a steepest descent type of update can be performed
	\begin{equation}
	\textbf{m}_{k+1} = \textbf{m}_k - \alpha\displaystyle\nabla \mathcal{L}\left(\textbf{m}\right)_k,
	\label{gbinv10}
	\end{equation}
where $\alpha$ is a step length parameter that is generally adapted dynamically. 

In this study, we used the Adam algorithm \citep[][]{Kingma-Ba2015} implemented in Pytorch to update $\textbf{z}_{k}$ using the $p\left(\textbf{z}\right) = U\left(\textbf{-1},\textbf{1}\right)$ prior directly. Adam is an advanced gradient-based algorithm that only makes use of the loss function gradient \citep[see][for details]{Kingma-Ba2015}. We used the following algorithmic parameters, a learning rate of 0.01 and the default values of 0.9 and 0.999 for the $\beta_1$ and $\beta_2$ momentum parameters. Since our considered forward operator, $F\left(\textbf{m}\right)$, is linear (see section \ref{meth_invprob}), it implies that for continuous $\textbf{m} = G\left(\textbf{z}\right)$ realizations it is straightforward to have the DL library internally computing $\nabla \mathcal{L}\left(\textbf{z}\right)$ (at least for common loss functions $\mathcal{L}\left(\cdot\right)$). Yet this is no longer possible if the $G\left(\textbf{z}\right)$ realizations require thresholding. To circumvent this issue, we implemented an adjoint model giving $\nabla \mathcal{L}\left(\textbf{m}\right)$ within the Pytorch computational graph \citep{Richardson2018}. Here $\nabla \mathcal{L}\left(\textbf{m}\right)$ is backpropagated through the generator network to update, using Adam, the latent vector, $\textbf{z}$, that produced the model, $\textbf{m} = G\left(\textbf{z}\right)$. As explained earlier, this is possible because the generator is fully differentiable end-to-end and $\nabla \mathcal{L}\left(\textbf{z}\right) = \nabla \mathcal{L}\left(\textbf{m}\right)\textbf{V}$  (see also our companion code for more details). 

A simple sum of squared errors (SSR, see equation \ref{gpr2}) was chosen for the loss function, $\mathcal{L}\left(\textbf{m}\right)$. Consequently, specific steps are needed to deal with the fact that the estimates in $\textbf{z}$ can move outside the support of the $U\left(\textbf{-1},\textbf{1}\right)$ prior. Three options to deal with this situation have been investigated by \citet{Lipton-Tripathi2017} in the context of direct inversion of a full (face) image: (i) allow $z$ values to leave the $\left[\textbf{-1},\textbf{1}\right]$ interval, (ii) replace components that are too large with the maximum allowed value and components that are too small with the minimum allowed value (i.e., standard clipping), (iii) reassign the exceeding $\textbf{z}$ components uniformly at random in the $\left[\textbf{-1},\textbf{1}\right]$ range (i.e., stochastic clipping). In agreement with the findings by \citet{Lipton-Tripathi2017}, limited testing revealed that stochastic clipping produces slightly better results for our considered case studies. Therefore, stochastic clipping is adopted throughout.

\subsection{Global Optimization}
\label{meth_globalopt}

For comparison purposes, we used the DREAM$_{\left(\rm ZS\right)}$ algorithm \citep[][]{Laloy-Vrugt2012,Vrugt2016} as global optimizer. This is a MCMC sampler designed to sample the posterior density function of the parameters. Yet it is herein used to solely find an appropriate solution rather than sampling the full posterior distribution. The DREAM$_{\left(\rm ZS\right)}$ scheme runs parallel interacting Markov chains and is thus population-based, and makes use of differential evolution principles to propose candidate points. It has been proven efficient in many hydrologic and geophysical applications \citep[see, e.g.,][for references]{Vrugt2016}. Full algorithmic details can be found in \citet{Laloy-Vrugt2012} and \citet{Vrugt2016}. Using a population-based search method requires finding the right trade-off between the size of the population (or ``exploration") and the number of generations (or ``exploitation") for a given number of forward model runs. After some testing, we selected 8 Markov chains, meaning a 8-member population, and performed as many MCMC iterations, that is, population generations, as necessary to fit the data to the assigned noise level.

\subsection{Inverse Problem}
\label{meth_invprob}

We consider a synthetic 2D straight-ray cross-hole GPR tomography exemple. Given that the physics is assumed to be fully linear, any convergence problems arising in solving the inverse problem can be attributed to the nonlinear relationship between $\textbf{z}$ and $G\left(\textbf{z}\right)$. We use two binary images with two facies of homogeneous GPR velocity (0.06 m ns$^{-1}$ and 0.08 m ns$^{-1}$) as true models (see Figures \ref{fig2}a, \ref{fig3}a, \ref{fig4}a, and \ref{fig5}a). Both are produced using $\textbf{m} = G\left(\textbf{z}\right)$ with $\textbf{z}$ randomly sampled from $U(\textbf{-1},\textbf{1})$. Cross-hole GPR imaging uses a transmitter antenna to emit a high-frequency electromagnetic wave at a location in one borehole and a receiver antenna to record the arriving energy at a location in another borehole \citep[e.g.,][]{Annan2005}. The considered measurement data are first-arrival traveltimes for several transmitter and receiver locations. These data contain information about the GPR velocity distribution between the boreholes. The GPR velocity primarily depends on dielectric permittivity, which is strongly influenced by volumetric water content and, consequently, porosity in saturated media. Our setup consists of two vertical boreholes that are located 6.0 m apart. Sources (left) and receivers (right) are located between 0.5 and 12.0 m depth with 0.5 m spacing (Figures \ref{fig2}a, \ref{fig3}a, \ref{fig4}a, and \ref{fig5}a), leading to a total dataset of $d_N$ = 576 traveltimes. Under the linear physics assumption, synthetic traveltime data, $\textbf{d}$, are simulated as
	\begin{equation}
	\textbf{d} = \textbf{A}\textbf{m} + \textbf{e},
	\label{gpr1}
	\end{equation}
where $\textbf{A} = \textbf{J}$ contains the path length in each model cell, $\textbf{e}$ represents independent random draws from a zero-mean homoscedastic Gaussian distribution with a typical standard deviation $\sigma_e$= 1 ns. When a simple SSR loss function (without regularization) is used
	\begin{equation}
	\mathcal{L}\left(\textbf{m}\right) = ||\textbf{d} - \textbf{A}\textbf{m}||^2 \equiv \mathcal{L}\left(\textbf{z}\right) = ||\textbf{d} - \textbf{A}G\left(\textbf{z}\right)||^2,
	\label{gpr2}
	\end{equation}
then the gradient vector of $\mathcal{L}\left(\textbf{m}\right)$ with respect to $\textbf{m}$ becomes
	\begin{equation}
	\nabla \mathcal{L}\left(\textbf{m}\right) = -2\textbf{A}^{\rm T \it}\left[\textbf{d} - \textbf{A}\textbf{m}\right]^{\rm T \it} \equiv \nabla \mathcal{L}\left(G\left(\textbf{z}\right)\right) = -2\textbf{A}^{\rm T \it}\left[\textbf{d} - \textbf{A}G\left(\textbf{z}\right)\right]^{\rm T \it}.
	\label{gpr3}
	\end{equation}
As explained above, this allows us to use Pytorch to backpropagate $\nabla \mathcal{L}\left(\textbf{m}\right)$ (equation (\ref{gpr3})) through the GAN generator and optimize $\textbf{z}$ such that equation (\ref{gpr2}) is minimized within a single computational graph using gradient descent. From now on, this will be referred to a strategy 1. In addition, estimating $\textbf{J}^{\rm z_{\rm SN}}_k$ at each iteration $k$ by a centered 2-point finite difference scheme together with using a GN search as described by equations \ref{gbinv5}-\ref{gbinv9} will be called strategy 2. Here $\textbf{J}^{\rm z_{\rm SN}}$ is derived using a perturbation factor of 0.1. This perturbation value was found to be the most appropriate after testing with several values in the $\left[0.01, 0.5\right]$ range. A perturbation of 0.1 for a $N\left(\textbf{0},\textbf{1}\right)$ prior might seem large. Yet one must keep in mind that the necessary thresholding of the produced realizations, $G\left(\textbf{z}\right)$, can cause one or more sensitivies in $\textbf{J}^{\rm z_{\rm SN}}$ to be zero or near-zero if the used perturbation is too small. Such excessively small  $J^{\rm z_{\rm SN}}_{i,j}$ values will induce instabilities in the GN search described by equation (\ref{gbinv9}).

Lastly, note that besides true models I and II we use two different realizations of $\textbf{e}$ in equation (\ref{gpr1}) which leads to two different ``true'' datasets for each true model. Realization I induces a true root mean square error (RMSE) between ``true'' (corrupted) and uncorrupted measurements of 1.006 ns. For realization II, the corresponding RMSE is 1.001 ns. Repeating the inversions with different noise realizations is useful to check the robustness of the inversion results against the exact noise values. Overall, this leads to 4 inverse case studies: 2 true models $\times$ 2 noise realizations. To ease the comparisons between the different inversion runs, we use the weighted root mean square error (WRMSE) which we define as the ratio of the achieved RMSE by the inversion to the true RMSE. A WRMSE $> 1$ ns thus basically means that the data are underfitted while WRMSE $= 1$ denotes an appropriate data misfit.

\section{Results}
\label{results} 

\subsection{Inverting the Full Image}
\label{res_fulliminv}

\begin{figure}[H]
	\noindent\hspace{-2.5cm}\includegraphics[width=45pc]{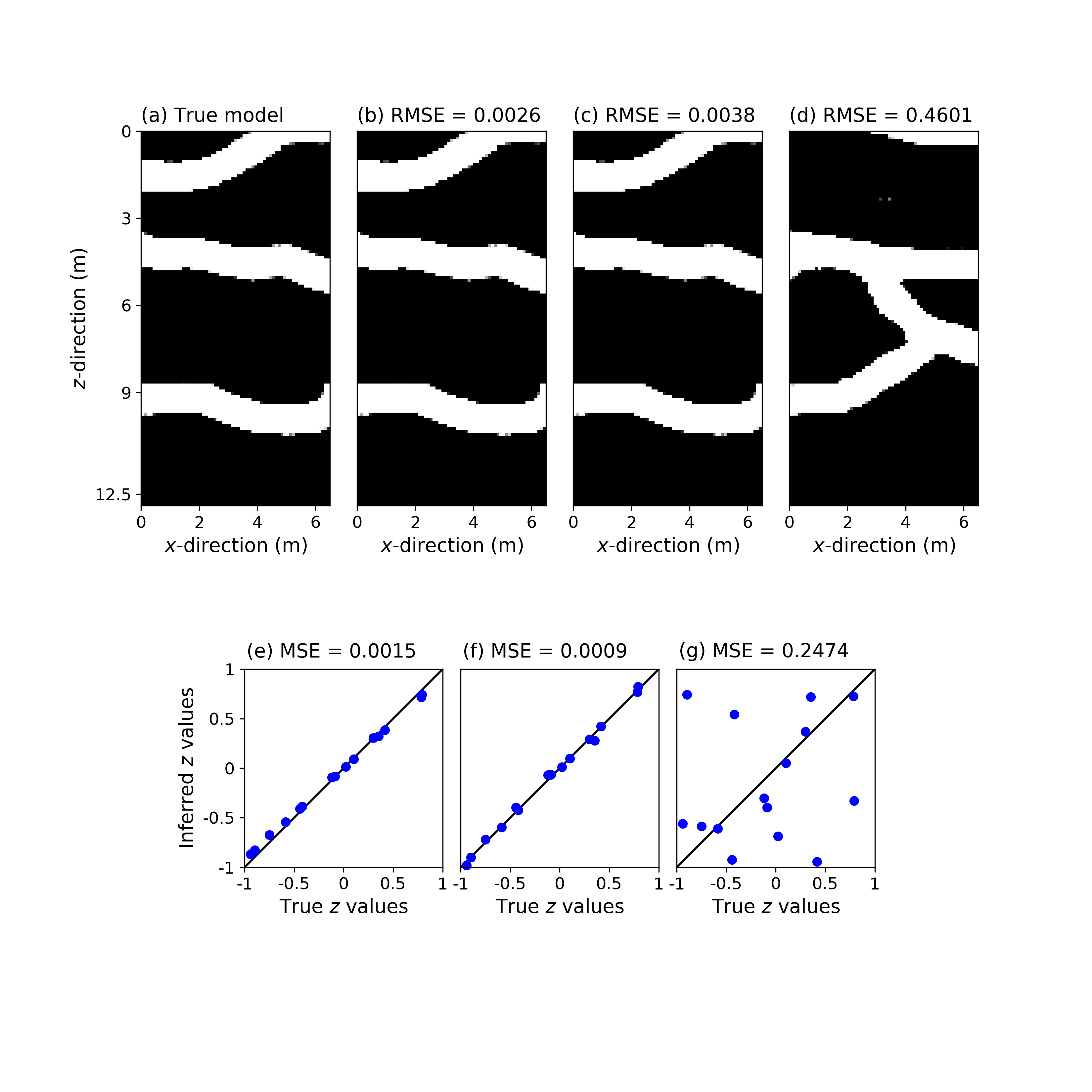}
	\vspace*{-22mm}
	\caption{Inversion results for the full image inversion using the true model I: (a) true model, (b) and (c): retrieved models with successful image recovery, (d) retrieved model with unsuccessful image recovery, (e - g) scatter plot of the recovered $\textbf{z}$ vector associated with the model shown in subplot (b - d) and the true $\textbf{z}$ vector, respectively.}
	\label{fig2}
\end{figure}

\begin{figure}[H]
	\noindent\hspace{-2.5cm}\includegraphics[width=45pc]{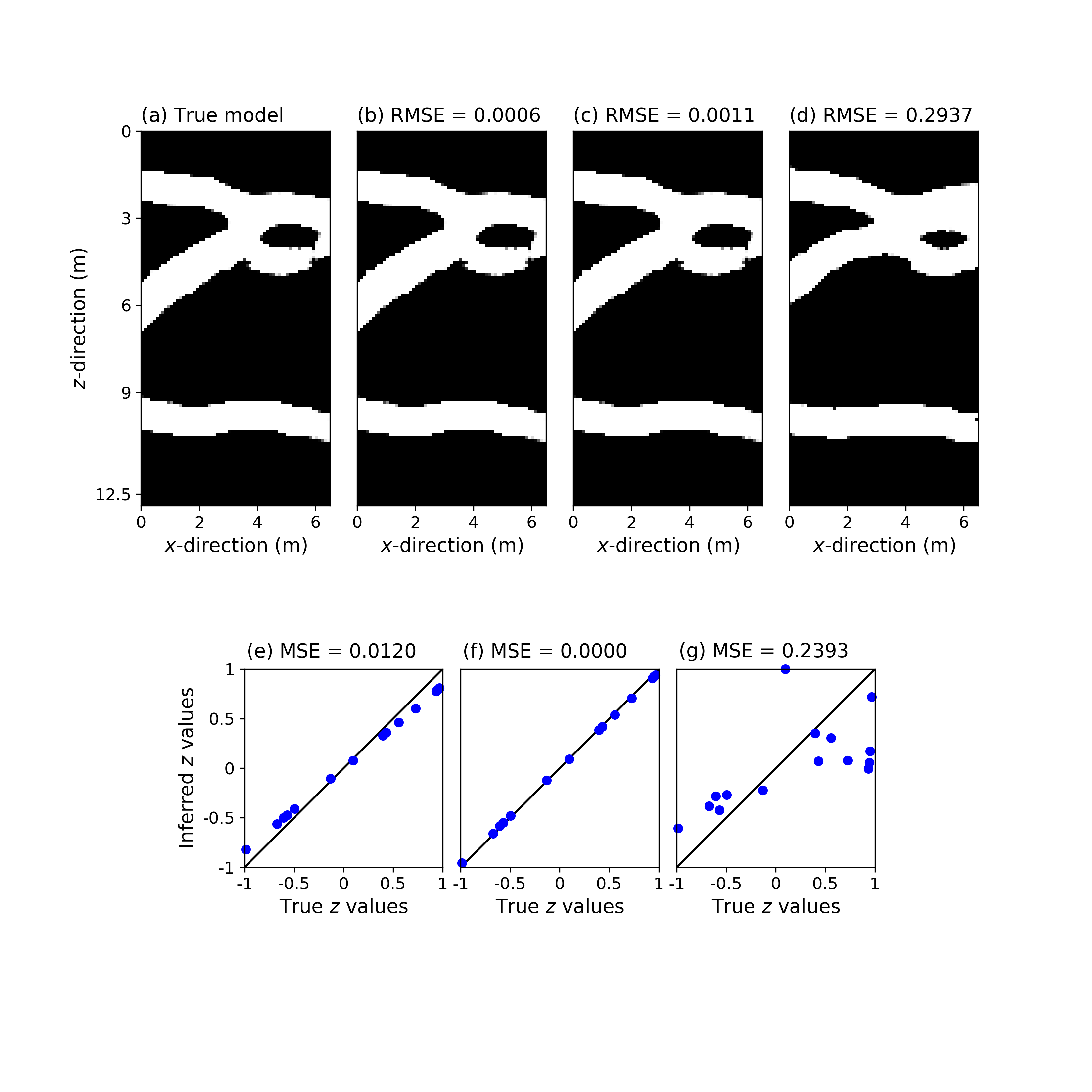}
	\vspace*{-22mm}
	\caption{Inversion results for the full image inversion using the true model II: (a) true model, (b) and (c): retrieved models with successful image recovery, (d) retrieved model with unsuccessful image recovery, (e - g) scatter plot of the recovered $\textbf{z}$ vector associated with the model shown in subplot (b - d) and the true $\textbf{z}$ vector, respectively.}
	\label{fig3}
\end{figure}

Before considering application of the forward operator, we tested whether gradient descent within the latent space of our trained GAN is able to recover the true (non-thresholded) $G\left(\textbf{z}\right)$ model and corresponding $\textbf{z}$ vector from the inversion of $G\left(\textbf{z}\right)$. In other words, is it possible to recover the values in $\textbf{z}$ given exhaustive knowledge of the true $G\left(\textbf{z}\right)$? This task is similar to the one considered by \citet{Lipton-Tripathi2017}, but in a geophysical context rather than for face images. The considered two true models in $\left[0,1\right]$ (I and II) generated by our trained GAN are depicted in Figures \ref{fig2}a and \ref{fig3}a. No data corruption was used and 50 inversion trials using 10,000 Adam iterations were performed with randomly chosen starting points. Table \ref{table1} and Figures \ref{fig2}-\ref{fig3} summarize the results. In Table \ref{table1}, a RMSE of 0.02 between the true and reconstructed images is selected to define a successfull reconstruction. This choice is based on the observation that, visually, the retrieved models are either very similar to the true images with a RMSE $<$ 0.02 or largely differ from the true images with a RMSE $>$ 0.2 - 0.3. Furthermore, similarly to \citet{Lipton-Tripathi2017}, we report the percentage of inversion results for which the mean squared error (MSE) between the true and inferred $z$ values is $<$ 0.01. Overall, 37\% of the inversion runs recover the true models (Table \ref{table1}). Furthermore, each of these successful runs corresponds to a successful recovery in the latent space for true model I (Table \ref{table1} and Figure \ref{fig2}) but less for model II (Table \ref{table1} and Figure \ref{fig3}). As illustrated by Figures \ref{fig3}b and \ref{fig3}e, this is because a slight linear bias in the $z$ values can nevertheless produce good image reconstructions. Our results somewhat differ from those obtained by \citet{Lipton-Tripathi2017} who report 100\% recovery in the $\textbf{z}$-space to arbitrary precision. There are two major differences between our work and that of \citet{Lipton-Tripathi2017}. First, they consider face images which are much more clustered compared to our channelized model: the eyes, nose, ears, etc. tend to occupy the same portion of the image and they are located at globally similar distances from each other. Second, \citet{Lipton-Tripathi2017} performed 100,000 gradient descent iterations while we performed 10 times less iterations.

\begin{table}[H]
	\caption{Percentage (\%) of inversion runs that achieved a RMSE $\leq$ 0.02 in recontructing the full 125 $\times$ 60 image ($G\left(\textbf{z}\right)$) and a MSE $\leq$ 0.01 in recovering the true 15-dimensional $\textbf{z}$ vector, for each true model (I and II).}
	\begin{adjustbox}{center}
		\begin{tabular}{cccccc}%
			\hline
			True model & RMSE $<$ 0.02 &  MSE $<$ 0.01 \\
			\hline
			1 & 34 & 32 \\
			2 & 40 & 8 \\
			\hline
		\end{tabular}
	\end{adjustbox}
	\label{table1}
\end{table}

\subsection{Geophysical Inversion}

We now turn our attention to the actual geophysical inversion, in which a subset of linear averages are observed under noise. For each strategy (1 and 2) and each combination of true model (I or II) and noise realization (I and II), the inversion was repeated 100 times using 100 different (randomly chosen) starting models. For strategy 1, the maximum total number of iterations (i.e., proposal steps) was set to 10,000. Here each Adam iteration consumes one forward solve. Since the resulting computational expense may be overly large for realistic nonlinear geophysical inverse problems, inversion results are also looked at after a total of 100 and 1000 iterations. Due to the used centered 2-point finite-difference approximation of $\textbf{J}^{\rm z_{\rm SN}}$ (see above), for strategy 2 one iteration incurs a cost of 31 forward solves. Therefore, the maximum number of iterations for strategy is set to 5000 which translates into 155,000 forward model runs. In addition, inversion performance is also considered after 100 (3100 forward solves) and 1000 (31,000 forward solves) iterations. Moreover, to verify the impact of thresholding on the quality of the inferred models, each case study is also run without thresholding. 

Strategy 1 without tresholding relies on the direct computation of $\nabla \mathcal{L}\left(\textbf{z}\right)$ by autodifferentiation with no adjoint model involved and is therefore basically the same as the approach by \citet{Bora2017}. For the continuous case, we consider the continuous versions of the true models shown in Figures  \ref{fig2}a and \ref{fig3}a and the same noise realizations as for the binary case. To convert a generated continuous model $\textbf{m} = G\left(\textbf{z}\right)$ in $\left[\textbf{0},\textbf{1}\right]$ into a continuous velocity field $\textbf{m}_{\rm VEL}$ (m ns$^{-1}$) in $\left[\textbf{0.06},\textbf{0.08}\right]$, the following relationship is used: $\textbf{m}_{\rm VEL} = 0.06 + 0.02\left(\textbf{1}-\textbf{m}\right)$ 

\begin{figure}[H]
	\noindent\hspace{-2.5cm}\includegraphics[width=45pc]{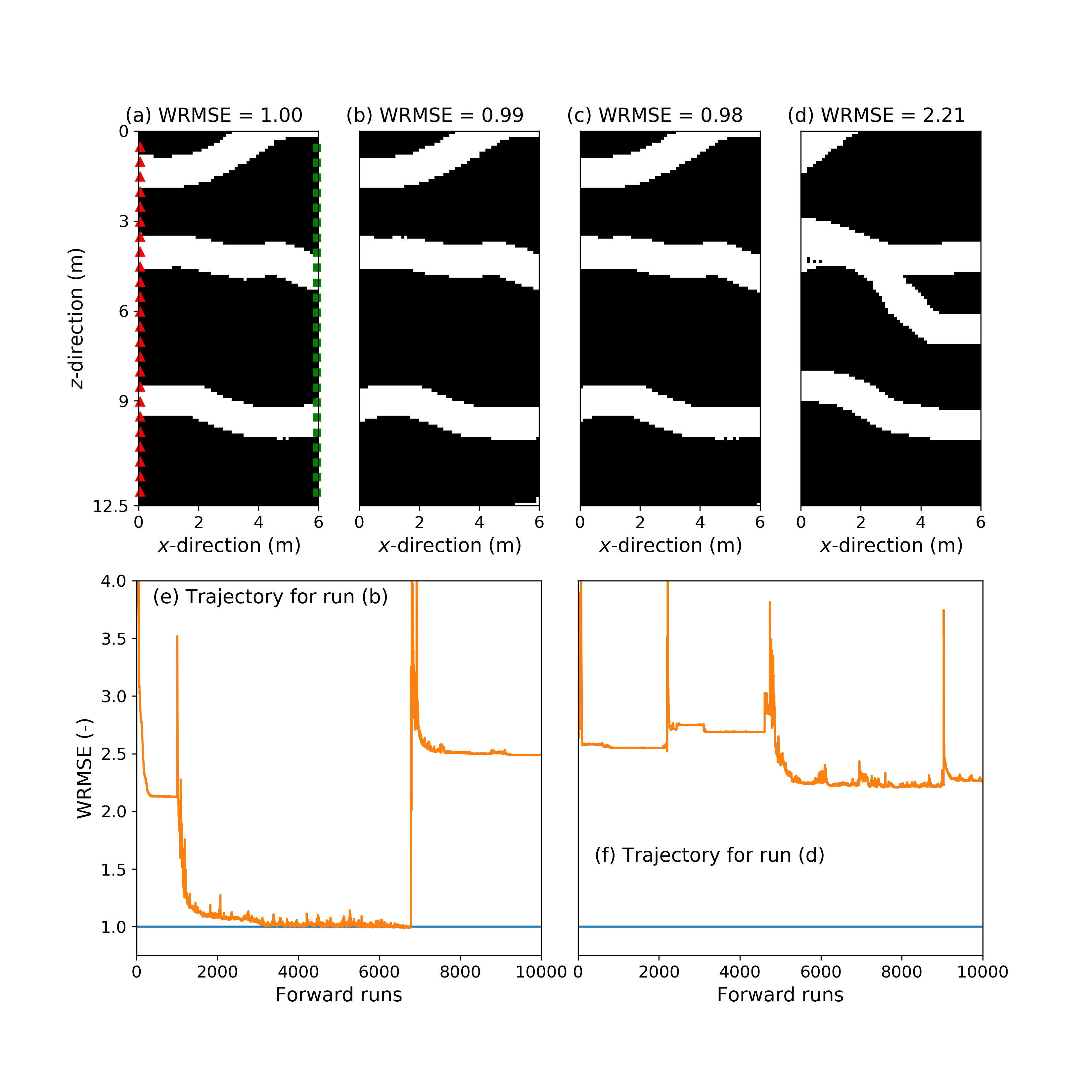}
	\vspace*{-22mm}
	\caption{Binary case: inversion results for strategy 1, truel model I and noise realization II: (a) true model, (b) and (c) two randomly chosen best-fitting models found over 100 repetitions, (d) best-fitting derived model across the allowed iterations of a randomly chosen repetition for which the corresponding best WRMSE is $> 1.2$ ns, (e) sampled WRMSE trajectory for the best model found among the 100 repetitions (displayed in subplot(b)), and (f) sampled WRMSE trajectory associated with the model depicted in subplot (d). The red triangles and the green squares in subfigures (a-d) represent the GPR source and receiver positions, respectively.}
	\label{fig4}
\end{figure}

\begin{figure}[H]
	\noindent\hspace{-2.5cm}\includegraphics[width=45pc]{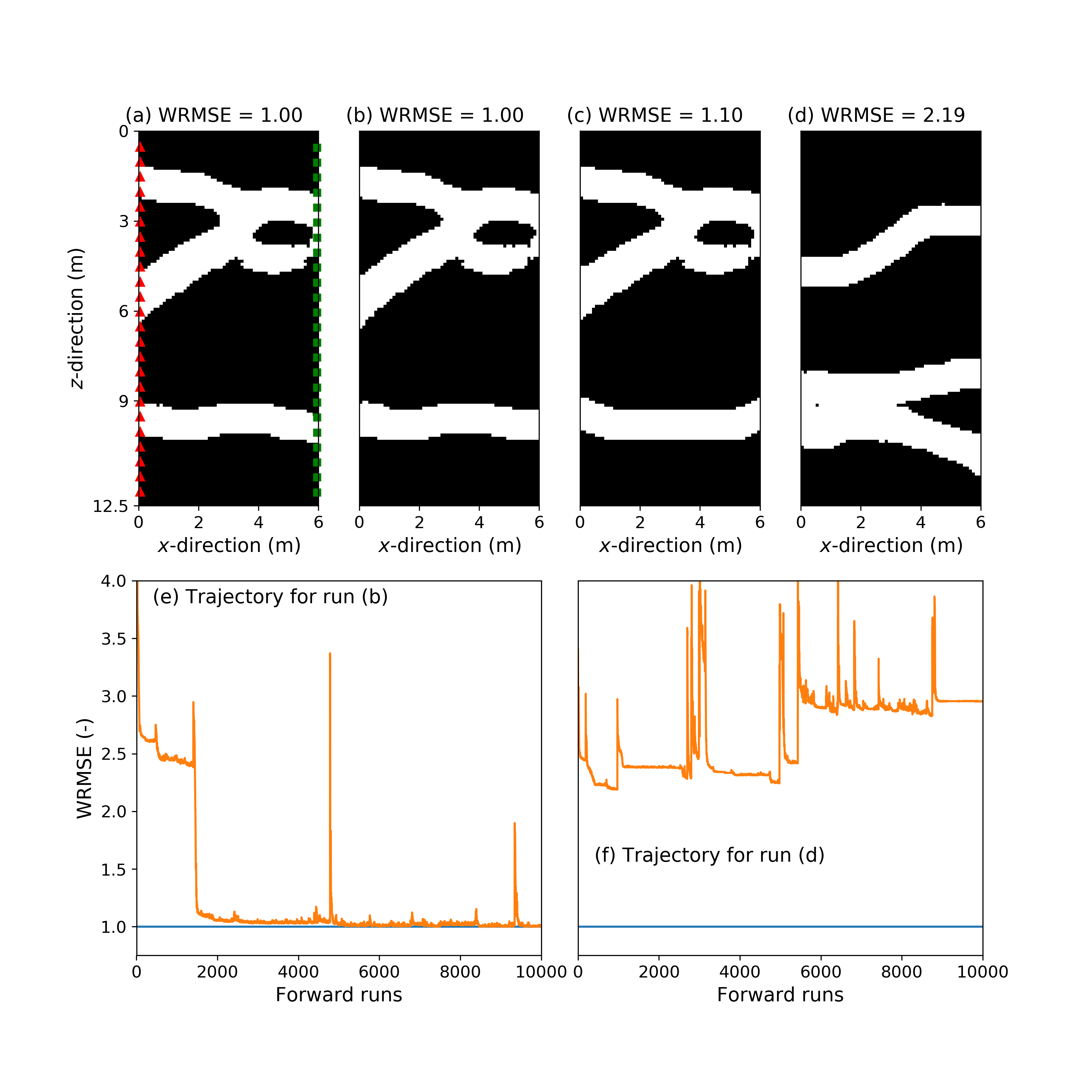}
	\vspace*{-22mm}
	\caption{Binary case: inversion results for strategy 1, truel model II and noise realization II: (a) true model, (b) and (c) two randomly chosen best-fitting models found over 100 repetitions, (d) best-fitting derived model across the allowed iterations of a randomly chosen repetition for which the corresponding best WRMSE is $> 1.2$ ns, (e) sampled WRMSE trajectory for the best model found among the 100 repetitions (displayed in subplot(b)), and (f) sampled WRMSE trajectory associated with the model depicted in subplot (d). The red triangles and the green squares in subfigures (a-d) represent the GPR source and receiver positions, respectively.}
	\label{fig5}
\end{figure}

\begin{figure}[H]
	\noindent\hspace{-2.5cm}\includegraphics[width=45pc]{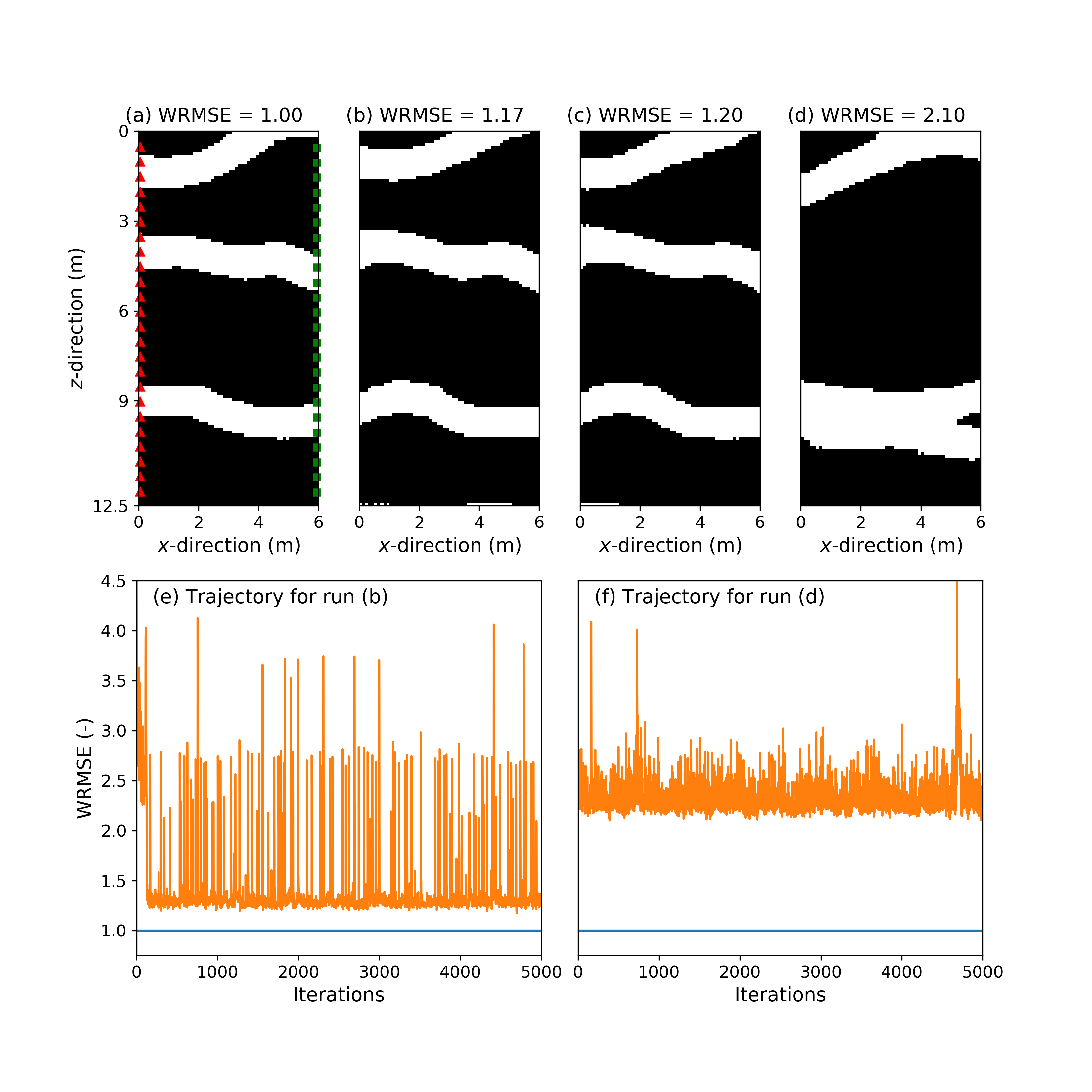}
	\vspace*{-22mm}
	\caption{Binary case: inversion results for strategy 2, truel model I and noise realization II: (a) true model, (b) and (c) two randomly chosen best-fitting models found over 100 repetitions, (d) best-fitting derived model across the allowed iterations of a randomly chosen repetition for which the corresponding best WRMSE is $> 1.2$ ns, (e) sampled WRMSE trajectory for the best model found among the 100 repetitions (displayed in subplot(b)), and (f) sampled WRMSE trajectory associated with the model depicted in subplot (d). The red triangles and the green squares in subfigures (a-d) represent the GPR source and receiver positions, respectively.}
	\label{fig6}
\end{figure}

\begin{figure}[H]
	\noindent\hspace{-2.5cm}\includegraphics[width=45pc]{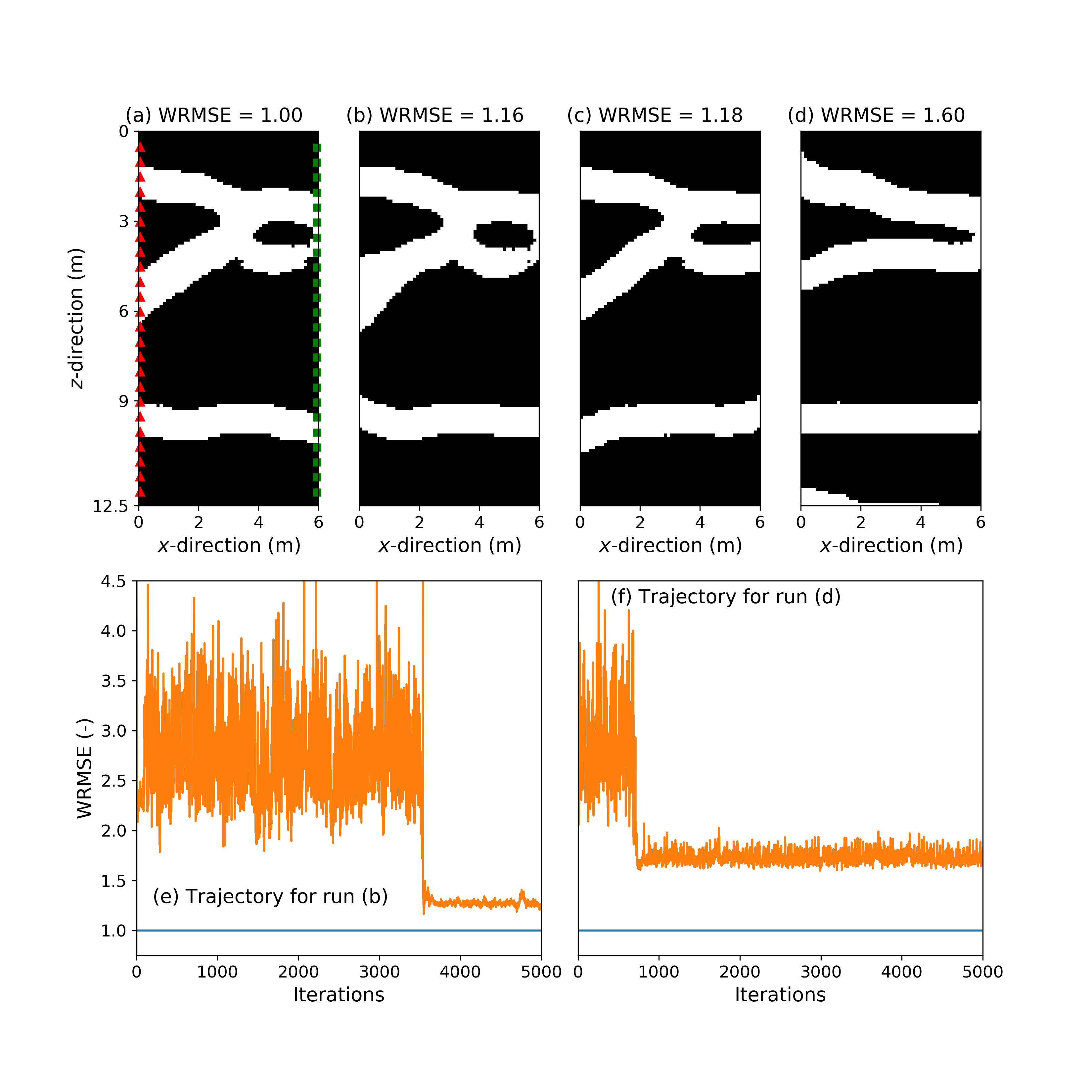}
	\vspace*{-22mm}
	\caption{Binary case: inversion results for strategy 2, truel model II and noise realization II: (a) true model, (b) and (c) two randomly chosen best-fitting models found over 100 repetitions, (d) best-fitting derived model across the allowed iterations of a randomly chosen repetition for which the corresponding best WRMSE is $> 1.2$ ns, (e) sampled WRMSE trajectory for the best model found among the 100 repetitions (displayed in subplot(b)), and (f) sampled WRMSE trajectory associated with the model depicted in subplot (d). The red triangles and the green squares in subfigures (a-d) represent the GPR source and receiver positions, respectively.}
	\label{fig7}
\end{figure}

The results of the 100 repetitions of the 8 different inversions are summarized in Table \ref{table2} for the continuous case and in Table \ref{table3} for the binary case. Furthermore, Figures \ref{fig4}-\ref{fig7} depict for each strategy applied to the binary case, true models I and II, and noise realization II, the true model (subplot (a)), two best-fitting models found among the 100 repetitions (subplots (b) and (c)), the best-fitting model across the allowed iterations of a randomly chosen trial for which the corresponding best WRMSE is $> 1.2$ ns (subplot (d)), the sampled WRMSE trajectory corresponding to the best-fitting model shown in subplot (b) (subplot (e)) and  the sampled WRMSE trajectory associated with the model displayed in subplot (d) (subplot (f)).

\begin{table}[H]
	\small
	\caption{Continuous case: number of inversion runs among the 100 performed ones that achieved a WRMSE $\leq$ 1.2, $\leq$ 1.1 and $\leq$ 1, respectively, for each inversion strategy (1 and 2), total number of iterations, N$_{\rm iter}$, combination of true model (I and II) and noise realization (I and II). The WRMSE is defined for each inversion run as the ratio of the best RMSE (over the iterations of a given run) to the RMSE of the true data (1.006 ns for noise realization I and 1.001 ns for noise realization II).}
	\begin{adjustbox}{center}
		\begin{tabular}{ccccccc}%
			\hline
			Strategy & N$_{\rm iter}$  & True model & Noise & WRMSE $\leq$ 1.2 & WRMSE $\leq$ 1.1 &  WRMSE $\leq$ 1.01 \\
			\hline
			1 & 100 & I & I & 0 & 0 & 0\\
			1 & 100 & I & II & 0 & 0 & 0\\
			1 & 100 & II & I & 0 & 0 & 0\\
			1 & 100 & II & II & 0 & 0 & 0\\
			1 & 1000 & I & I & 3 & 3 & 0\\
			1 & 1000 & I & II & 6 & 1 & 0\\
			1 & 1000 & II & I & 3 & 1 & 0\\
			1 & 1000 & II & II & 3 & 2 & 0\\
			1 & 10,000 & I & I & 24 & 21 & 16\\
			1 & 10,000 & I & II & 25 & 24 & 22\\
			1 & 10,000 & II & I & 14 & 13 & 4\\
			1 & 10,000 & II & II & 20 & 16 & 9\\
			2 & 100 & I & I & 19 & 0 & 0\\
			2 & 100 & I & II & 6 & 2 & 0\\
			2 & 100 & II & I & 0 & 0 & 0\\
			2 & 100 & II & II & 1 & 0 & 0\\
			2 & 1000 & I & I & 100 & 1 & 0\\
			2 & 1000 & I & II & 84 & 8 & 0\\
			2 & 1000 & II & I & 0 & 0 & 0\\
			2 & 1000 & II & II & 2 & 1 & 0\\
			2 & 5000 & I & I & 100 & 6 & 0\\
			2 & 5000 & I & II & 100 & 14 & 0\\
			2 & 5000 & II & I & 2 & 0 & 0\\
			2 & 5000 & II & II & 2 & 2 & 0\\
			\hline
		\end{tabular}
	\end{adjustbox}
	\label{table2}
\end{table}
\begin{table}[H]
	\small
	\caption{Binary case: number of inversion runs among the 100 performed ones for the continuous case that achieved a WRMSE $\leq$ 1.2, $\leq$ 1.1 and $\leq$ 1, respectively, for each inversion strategy (1 and 2), total number of iterations, N$_{\rm iter}$, combination of true model (I and II) and noise realization (I and II). The WRMSE is defined for each inversion run as the ratio of the best RMSE (over the iterations of a given run) to the RMSE of the true data (1.006 ns for noise realization I and 1.001 ns for noise realization II).}
	\begin{adjustbox}{center}
		\begin{tabular}{ccccccc}%
			\hline
			Strategy & N$_{\rm iter}$  & True model & Noise & WRMSE $\leq$ 1.2 & WRMSE $\leq$ 1.1 &  WRMSE $\leq$ 1.01 \\
			\hline
			1 & 100 & I & I & 0 & 0 & 0\\
			1 & 100 & I & II & 1 & 0 & 0\\
			1 & 100 & II & I & 0 & 0 & 0\\
			1 & 100 & II & II & 0 & 0 & 0\\
			1 & 1000 & I & I & 5 & 3 & 0\\
			1 & 1000 & I & II & 4 & 3 & 0\\
			1 & 1000 & II & I & 3 & 2 & 0\\
			1 & 1000 & II & II & 1 & 0 & 0\\
			1 & 10,000 & I & I & 21 & 19 & 13\\
			1 & 10,000 & I & II & 24 & 23 & 19\\
			1 & 10,000 & II & I & 10 & 9 & 7\\
			1 & 10,000 & II & II & 24 & 17 & 12\\
			2 & 100 & I & I & 22 & 0 & 0\\
			2 & 100 & I & II & 8 & 3 & 0\\
			2 & 100 & II & I & 0 & 0 & 0\\
			2 & 100 & II & II & 0 & 0 & 0\\
			2 & 1000 & I & I & 100 & 0 & 0\\
			2 & 1000 & I & II & 57 & 4 & 0\\
			2 & 1000 & II & I & 3 & 0 & 0\\
			2 & 1000 & II & II & 2 & 0 & 0\\
			2 & 5000 & I & I & 100 & 0 & 0\\
			2 & 5000 & I & II & 97 & 5 & 0\\
			2 & 5000 & II & I & 6 & 0 & 0\\
			2 & 5000 & II & II & 9 & 1 & 0\\
			\hline
		\end{tabular}
	\end{adjustbox}
	\label{table3}
\end{table}

It is observed for both strategies 1 and 2, that the quality of the best-fitting solution heavily depends on the starting model and the allowed number of iterations. Indeed for the smallest budget of 100 iterations, no satisfying solutions are found no matter the inversion approach. We find that strategy 1 (Figures \ref{fig4}-\ref{fig5} and Tables \ref{table2}-\ref{table3}) shows a much higher success rate than strategy 2 (Figures \ref{fig6}-\ref{fig7} and Tables \ref{table2}-\ref{table3}) for the largest number of iterations allowed to each method. Hence strategy 1 retrieves solutions with a WRMSE $\leq$ 1.01 in 4\% to 22\% of the trials for the continuous case and in 7\% to 19\% of the trials for the binary case. On the contrary, strategy 2 never does so (Tables \ref{table2}-\ref{table3}). 

Strikingly, there appears to be very little negative impact of thresholding (compare Table \ref{table2} to Table \ref{table3}), except for strategy 2 and true model I (compare the bottom row of Table \ref{table2} with those of Table \ref{table3}). This indicates that plugging the externally computed  $\nabla \mathcal{L}\left(\textbf{m}\right)$ into the DL graph to permit gradient descent works well for the considered examples.

When the maximum number of iterations is reduced to 1000, the proportion of solutions associated with a WRMSE $\leq$ 1.1 becomes small but is overall two times larger for strategy 1 (2\% on average over the 4 datasets) than for strategy 2 (1\% on average over the 4 datasets). Since a model with a WRMSE as large as 1.2 may still visually ressemble the true model well (see Figure \ref{fig6}c), it is instructive to also consider a WRMSE $\leq$ 1.2 as threshold. Here strategy 2 becomes more attractive, but for true model I only (Tables \ref{table2}-\ref{table3}). We have tried to improve strategy 2 by (i) damping the model update at the early iteration before increasingly adding more contrast to the proposed models and (ii) decreasingly smooth the proposed models after each iteration as the search progresses. Nevertheless, neither of those schemes proved successful.

There is a systematic difference in peformance between the cases with true models I and II, with true model I always inducing a larger success rate no matter the inversion strategy and computational effort (Tables \ref{table2} and \ref{table3}). This is due to the fact that model I has higher prior probability or, to use deep learning terminology, is closer to the range of the generator \citep[][]{Bora2017} than model II. This can be informally seen by comparing models I (Figure \ref{fig2}a) and II (Figure \ref{fig3}a) with the original TI (Figure \ref{fig1}a): the patterns displayed by model I are occuring more frequently in the TI than the eye-type of channel branching contained in model II. It is seen that strategy 1 offers greater robustness than strategy 2 against variations in prior model probability, as shown by the smaller differences in success rates between the model I and II cases for strategy 1 (Tables \ref{table2} and \ref{table3}). 

With respect to sampling trajectories of the inversions, for strategy 1 most of the data misift reduction in a productive run (i.e., a run with a favorable starting model) is achieved within the first 3000-5000 iterations (Figures \ref{fig4}e and \ref{fig5}e). The main reduction in data misfit for strategy 2 can occur both rather quickly ($<$ 500 iterations) or much later (Figures \ref{fig6}e and \ref{fig7}e), while this strategy hardly recovers models with WRMSE $\leq$ 1.1. In addition, the sampling behavior of both algorithms can be unstable (Figures \ref{fig4}e, \ref{fig5}e, \ref{fig6}e and \ref{fig7}e). In accordance with those findings, it is also noted for strategy 1 that if WRMSE values larger than 2-3 are still sampled after some 2000 forward simulations, then the considered run will be unable to obtain a high-quality solution (Figures \ref{fig4}f and \ref{fig5}f).

It is worth noting that using strategy 2 with a trained GAN for which the latent space obeys a standard normal distribution, $N\left(\textbf{0},\textbf{1}\right)$, thereby removing the additional nonlinearity caused by the $\Psi\left(\textbf{z}_{\rm SN \it, k}\right)$ transform, leads to globally similar results as those presented in Tables \ref{table2} and \ref{table3} for strategy 2.

\begin{figure}[H]
	\noindent\hspace{-2.3cm}\includegraphics[width=45pc]{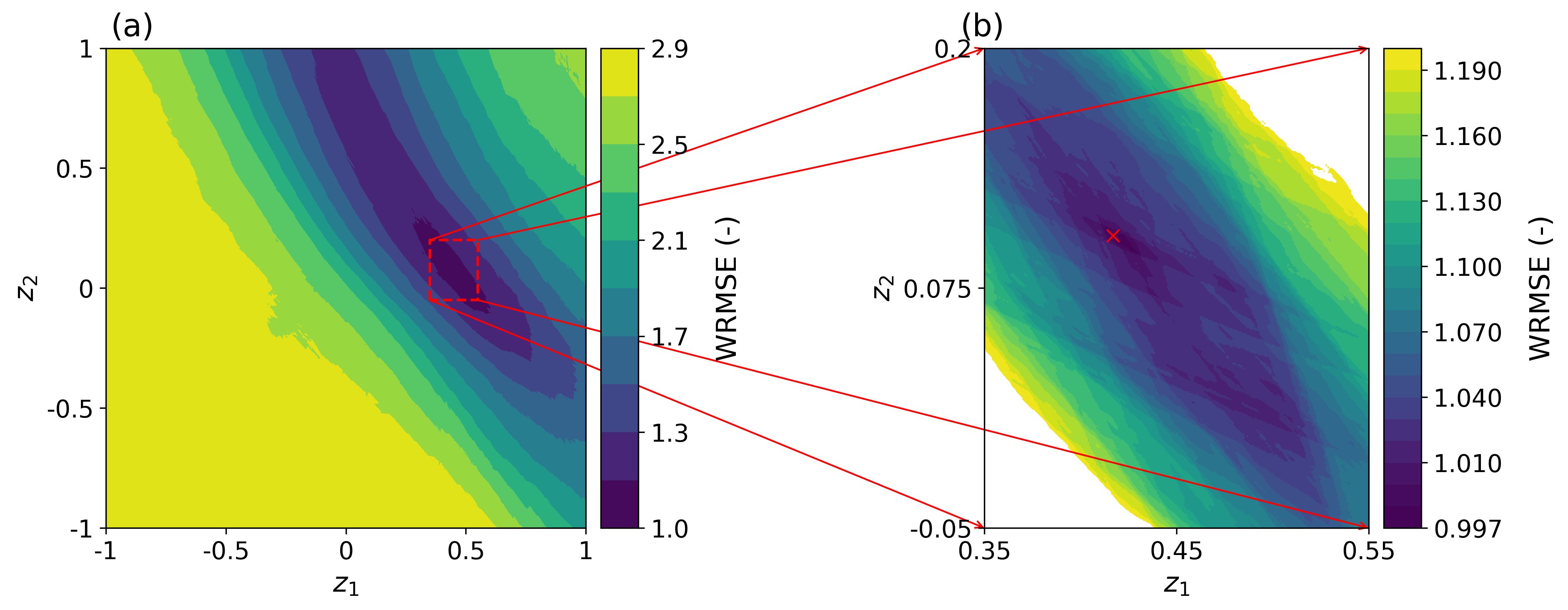}
	\caption{2D slice in the WRMSE landscape corresponding to the true reference solution obtained by the combination of the true model I (see Figure \ref{fig2}a, \ref{fig4}a or \ref{fig6}a) and noise realization II. The (a) subplot shows a coarse scale resolution over the full WRMSE range while the (b) subplot presents a zoom into the main basin of attraction. The red cross in the (b) subplot denotes the true values of $z_1$ and $z_2$ in this case.}
	\label{fig8}
\end{figure}

Even for our best performing strategy 1 and considered maximum number of iterations of 10,000, the overall success (WRMSE $\leq$ 1.01) rate remains relatively low (Tables \ref{table2} and \ref{table3}). As posited earlier in this paper, we argue that the reason for this is the high degree of nonlinearity associated with the relationship between $\textbf{z}$ and $G\left(\textbf{z}\right)$. More insights into the effect of the nonlinear $G\left(\textbf{z}\right)$ transform on the considered simple inverse problem are provided by Figure \ref{fig8}. The latter displays a 2D slice in the WRMSE landscape corresponding to the true reference solution obtained by the combination of the true model I (see Figure \ref{fig2}a, \ref{fig4}a or \ref{fig6}a) and noise realization II. To construct Figure \ref{fig8}, the $z_3$ to $z_{15}$ components of the true $\textbf{z}$ vector that induces model I are kept fixed while the $z_1$ and $z_2$ dimensions are gradually varied between -1 and 1. While the WRMSE response surface appears smooth when considered at a coarse resolution (Figure \ref{fig8}a), a zoom into the basin of attraction (Figure \ref{fig8}b) reveals that the problem is in fact multimodal with many local minima. Also, this WRMSE landscape contains many small spikes and some large flat areas.

Lastly, we find that performing a global optimization with DREAM$_{\left(\rm ZS\right)}$ parameterized as described in section \ref{meth_globalopt} provides a solution with a WRMSE of 1.000 for all of the 4 scenarios (true models I and II combined with noise realizations I and II). Nevertheless, this is at the cost of a serial total of 200,000 to 800,000 forward runs. Among these amounts, some 50,000 serial forward calls are consumed to reach a WRMSE of about 1.1 while the rest is spent by moving from 1.1 to $\leq$ 1. This points out that traveling across the relatively rough misfit terrain caused by the nonlinearities in $G\left(\textbf{z}\right)$ may require global optimization or search methods even in the best case of a linear forward problem. Although not done herein, note also that population-based methods such as DREAM$_{\left(\rm ZS\right)}$ are straightforward to evolve in parallel using one CPU per population member, which can save a large amount of time \citep[][]{Laloy2018}.

\section{Discussion}
\label{discussion}

This paper demonstrates that even if the forward model is linear, the high degree of nonlinearity in the relationship between latent vector, $\textbf{z}$, and the generator of a GAN, $G\left(\textbf{z}\right)$, can cause gradient-based deterministic inversion performed in the GAN latent space $p\left(\textbf{z}\right)$ to fail. This is here illustrated for a classical binary channelized aquifer TI. Less structured and/or continuous TIs might be more amenable to gradient-based deterministic inversion within $p\left(\textbf{z}\right)$ since the $G\left(\textbf{z}\right)$ transform might then be less nonlinear. The poor convergence to the target misfit is in stark contrast to the results obtained when inverting the full image without noise (Table \ref{table2}, Figures \ref{fig2}-\ref{fig3}). This suggests that limited sampling of spatial averages under noise further complicates the recovery of the true $z$ values.

Our findings might appear to contradict those of \citet{Bora2017} who investigated whether gradient descent within the latent space of a trained GAN can recover an image from a set of measurements obtained by applying a linear operator to the true image. In cases where this true image is itself created by running the trained GAN for a given $\textbf{z}$ vector (similarly as in our study), \citet{Bora2017} systematically find small reconstruction errors. Nonetheless, Bora and coworkers consider face images (from the Celeba dataset) which are thus very different than our channelized aquifer training image. Face images are comparatively more clustered as the same structures (nose, ears, eyes,...) are more or less always located in the same areas of the image. Yet the channels in our training set can be anywhere \citep[see][for the full 2500 $\times$ 2500 training image from which the training set is sampled]{Laloy2018}. This larger degree of freedom in our images (for this aspect) might make the inversion problem harder to solve. In addition, \citet{Bora2017} do not give much details about their exact experimental setup which makes an in-depth comparison with our results difficult.

It might seem surprising that, if a sufficient number of iterations is allowed, strategy 1, which is based on the objective function gradient, $\nabla \mathcal{L}\left(\textbf{m}\right)$, outperforms a GN search that uses a finite-difference approximation of the full Jacobian matrix in the $p\left(\textbf{z}\right)$ space, $\textbf{J}^{\rm z}$. In the training of deep neural networks, stochastic gradient descent has proven highly successful. \citet{Goodfellow2016} argue that such first-order methods may more easily escape saddle points, while second-order methods tend to get attracted and stuck in them. The situation might be similar here where we try to estimate $\textbf{z}$ for an already trained network, namely that the less ``stable" gradient-based methods can more easily move away from saddle-points and possibly local minima. Furthermore, strategy 1 relies on a rather powerful stochastic gradient descent algorithm \citep[Adam,][]{Kingma-Ba2015} and uses a perfectly accurate estimation of $\nabla \mathcal{L}\left(\textbf{m}\right)$ since in the considered case, $\nabla \mathcal{L}\left(\textbf{m}\right)$ is available analytically. On the contrary, when approximated by finite differencing $\textbf{J}^{\rm z}$ is inevitably fraught with errors. 

Note that for the gradient descent used with strategy 1, no optimization of the Adam's hyperparameters, that is, the learning rate, and $\beta_1$ and $\beta_2$ momentum parameters \citep[see][]{Kingma-Ba2015}, was done. Only the learning rates of 0.1 and 0.001 were found to lead to a worse performance than using our choice of 0.01. Other Adam's hyperparameter settings or other descent algorithms such as SGD, RMSPROP \citep[see, e.g.,][]{Goodfellow2016} or Adabound \citep[][]{Luo2019} might enabled improved inversion quality for the considered case studies.

Lastly, notice that some recent work suggests that projecting the learned latent space onto a so-called ``Riemannian"  manifold may lead to a reduced non-linearity of $G\left(\textbf{z}\right)$ with respect to the projected $\textbf{z}$ \citep[][]{Shao2017,Arvanitidis2018,Chen2018}. In the future, it might be worthwhile to attempt inversion within such a manifold.

\section{Concluding remarks}
\label{conclusion}

Performing inversion within the low-dimensional, latent space of a trained GAN, $p\left(\textbf{z}\right)$, rather than within the original model space, $p\left(\textbf{m}\right)$, offers important advantages in that (1) any generated model proposal honors the prior training image (TI) and (2) the parameter space is reduced by orders of magnitude compared to the original model space. Global probabilistic inversion within $p\left(\textbf{z}\right)$ has been recently shown to work well \citep[][]{Laloy2018}. However, such probabilistic inversion still incurs a large computational cost while the learned $p\left(\textbf{z}\right)$ could also be possibly used within computationally less demanding gradient-based deterministic inversions. In this work, we show that owing to the highly nonlinear relationship between the GAN generator, $G\left(\textbf{z}\right)$, and latent vector, $\textbf{z}$, gradient-based deterministic inversion may fail even though the physics of the forward problem is linear. For a channelized aquifer binary TI and a synthetic linear GPR tomography problem involving 576 measurements with low noise, we find that when allowing for a total of 10,000 iterations, about 13\% of the trials by the considered gradient descent algorithm locate a solution that has the required data misfit, in average over the considered test cases. When restricting the maximum allowed number of iterations to 1000, the percentage of success becomes 0\%, though approximately 2 \% of the inferred models still closely ressemble their associated true model and induce a nearly correct data misfit. Furthermore, the tested Gauss-Newton inversion approach that approximates the Jacobian matrix to create updates proved to be unable to recover a solution with the appropriate data misfit. Overall, deterministic inversion performance is found to depend on the inversion approach, starting model, true reference model, number of iterations and noise realization. In contrast, costly global optimization with DREAM$_{\left(\rm ZS\right)}$ always finds an appropriate solution.

\section{Computer Code Availability}
The GAN and associated inversion codes used in this study are available at \url{https://github.com/elaloy/gan_for_gradient_based_inv}.

\section{Used GAN architecture}
\label{appendix_archi}
Compared to the generator architecture detailed in \citet{Laloy2018}, we have the following differences.
\begin{itemize}
	\item Batch normalization was replaced by instance normalization
	\item The nonlinearity in the 5$^{\rm th}$ layer of the generator is a ReLU instead of an hyperbolic tangent.
	\item A first transposed dilated convolutional layer takes the output of the 5$^{\rm th}$ layer as input. This 6$^{\rm th}$ layer is parameterized with 64 output channels, a kernel size of 5 pixels, a stride of 1 pixel, a padding of 6 pixels, an output padding of 0 pixels and dilation coefficient of 3. The activation function in this 6$^{\rm th}$ layer is a ReLU and instance normalization is applied after the nonlinearity.
	\item A second transposed dilated convolutional layer takes the output of the 6$^{\rm th}$ layer as input. This 7$^{\rm th}$ layer is parameterized with 1 output channel, a kernel size of 5 pixels, a stride of 1 pixel, a padding of 10 pixels, an output padding of 0 pixels and dilation coefficient of 5. The nonlinearity in this 7$^{\rm th}$ layer is a hyperbolic tangent.
\end{itemize}

\end{document}